# Enhancing Mechanical Stimuli in Functionally Graded Bone Scaffolds Through Porosity Gradients: A Finite Element Analysis Study

Anson Wen Han Cheong [1], Vahid Badali [2,3], Sean Kiely [4], Iman Roohani [4], Yang Jiang [5], Jianguang Fang [5], Ali Entezari* [4]

[1] School of Mechanical and Manufacturing Engineering, University of New South Wales, Sydney, NSW 2052, Australia
[2] Department of Structural Mechanics and Analysis, Technische Universität Berlin, Berlin, Germany
[3] Julius Wolff Institute, Berlin Institute of Health, Charité—Universitätsmedizin Berlin, Berlin, Germany
[4] School of Biomedical Engineering, Faculty of Engineering and Information Technology, University of Technology Sydney, NSW 2007, Australia
[5] School of Civil and Environmental Engineering, University of Technology Sydney, Sydney, NSW 2007, Australia

* Corresponding Author
Email: ali.entezari@uts.edu.au

## Abstract

Achieving an optimal biomechanical environment within bone scaffolds is critical for promoting tissue regeneration, particularly in load-bearing anatomical sites where rigid fixation can induce stress shielding and compromise healing. Functionally graded (FG) scaffolds, which incorporate controlled variations in porosity or material properties, have attracted significant attention as a strategy to mitigate stress shielding by promoting more favourable load transfer. In this study, the effects of porosity gradient magnitude (i.e., max-to-min ratio of porosity), gradient resolution, scaffold material properties, and fixation plate rigidity on the distribution of mechanical stimuli within FG scaffolds were systematically investigated. Finite element analyses (FEA) were conducted on a femoral segmental defect model stabilised with a bone plate, and multiple porosity gradient strategies were compared against a corresponding uniform scaffold composed of body-centred cubic (BCC) unit cells. Scaffolds composed of titanium alloy (Ti-6Al-4V), bioactive glass (45S5 Bio-glass), and polylactic acid (PLA) were evaluated to capture a range of material stiffnesses. Introducing porosity gradients consistently enhanced the mean octahedral shear strain within the scaffold, particularly in regions adjacent to the fixation plate affected by stress shielding. The magnitude of mechanical stimulus improvement increased with both greater porosity gradient magnitudes and higher gradient resolution. These improvements were more pronounced in stiffer materials, such as Ti-6Al-4V, emphasising the critical interplay between scaffold material properties and architectural design. These findings highlight the importance of tailoring both porosity profiles and material selection to optimise scaffold mechanics for bone regeneration.

**1. Introduction**

Scaffold-based bone tissue engineering offers a promising strategy for repairing large bone defects caused by trauma, tumours, or degenerative diseases [1]. These scaffolds, typically made from biocompatible and osteoconductive materials, provide mechanical support and act as templates for new tissue formation [2]. Scaffolds also guide cell migration and extracellular matrix deposition, which are essential in bone regeneration and promoting patient recovery [3, 4].

Despite significant advancements in the development of bone tissue scaffolds, several challenges remain, hindering their widespread clinical translation. A key challenge for bone scaffolds is achieving an optimal balance between mechanical function and biological response, particularly for load-bearing applications [5]. These bone scaffolds are required to possess sufficient mechanical properties to withstand physiological loads, while also permitting adequate mechanical stimuli. Physical forces, such as the mechanical strain experienced by cells, promote cellular activity and bone tissue formation [6-8]. These mechanical cues are known to influence cell behaviour, particularly cellular differentiation, which is critical for effective bone regeneration and remodelling [4, 9-12]. Nevertheless, excessive scaffold stiffness or rigid fixation systems can induce stress shielding effects, reducing the mechanical stimuli transmitted to the regenerating tissue and thereby impairing the bone healing process [13].

Mechanical stimuli play a critical role in bone regeneration; its effects and optimisation have been outlined in several recent *in vivo* studies [14, 15]. Schouman et al. showed that reducing scaffold stiffness in porous titanium implants enhances bone ingrowth, driven by increased local strains that promote osteogenic differentiation [16]. Further demonstrated by Reznikov et al., bone ingrowth within scaffold pores is inversely related to scaffold stiffness, with more compliant scaffolds promoting greater strain-driven osteogenesis [17]. This becomes especially relevant when scaffolds are stabilised with fixation plates, where plate stiffness can induce stress shielding, resulting in non-uniform strain distribution within the scaffold. Studies have shown that scaffolds implanted in segmental defects with rigid fixations often experience inadequate or uneven mechanical loading, leading to suboptimal bone formation [18, 19]. These findings underscore the need for mechanically optimised scaffold designs.

Several factors influence the mechanical stimuli experienced within scaffolds, including physiological loading conditions, scaffold material properties, fixation constructs, and architectural features such as porosity, pore size, and shape [20-24]. While certain factors may be difficult to control in clinical settings, others, such as material composition and scaffold architecture, can be systematically optimised as design parameters [25]. By precisely tuning these properties, the mechanical environment within the scaffold can be modulated, highlighting the critical role of scaffold design in promoting osteogenic responses [26].

Extensive research has been dedicated to optimising scaffolds to enhance mechanical stimulation [27, 28]. Computational approaches, particularly finite element analysis (FEA) and topology optimisation, have demonstrated significant potential in designing scaffolds that provide optimal mechanical cues, ultimately improving bone regeneration [27, 29-32]. For instance, C. Wu et al. developed a dynamic mechanobiological topology optimisation framework to refine scaffold designs and promote sustained bone tissue formation in the long term [33]. While such studies have effectively improved mechanical stimulation in tissue scaffolds, they have largely focused on scaffolds with uniform architecture throughout the geometry, which may limit their ability to address spatially varying mechanical and biological demands across large defect sites.

Functionally graded (FG) scaffolds have emerged as a promising approach to overcome the limitations of uniform scaffold designs by introducing spatial variations in material properties and geometric features [34, 35]. Inspired by the hierarchical structure of natural tissues, FG scaffolds incorporate controlled gradients in porosity, stiffness, or composition, enabling local tailoring of mechanical and biological responses [36-38]. Unlike conventional scaffolds based on repeating uniform unit cells, FG scaffolds offer spatially optimised properties across the scaffold geometry, potentially improving mechanical efficiency, enhancing biological integration, and better accommodating the evolving physiological demands during bone regeneration [37].

Developing upon existing research, several studies have investigated how functional grading influences scaffold performance and have proposed various strategies for their design optimisation [39, 40]. Mehboob and Chang used a computational model of a composite bone plate–scaffold assembly to demonstrate that scaffolds with longitudinally graded material stiffness can achieve a more favourable balance between mechanical stability and biological healing compared to those with homogeneous stiffness [41]. Another recent example by Wu et al. developed a machine learning-based dynamic optimisation framework to design FG ceramic scaffolds with lateral and vertical porosity gradients, demonstrating that both grading strategies improved long-term bone regeneration compared to uniform scaffolds, with lateral gradients achieving superior outcomes [42]. In a related study, Wu et al. combined Bayesian optimisation and ceramic 3D printing to fabricate FG triply periodic minimal surface (TPMS) scaffolds, demonstrating improved bone ingrowth by tailoring scaffold architecture according to subject-specific loading conditions [43]. Similarly, Boccaccio et al. developed a mechanobiology-based optimisation algorithm to design scaffold microstructures that maximise bone formation based on mechanical stimuli [44]. By integrating FEA with a mechano-regulation model, they demonstrated that scaffold geometry and stiffness significantly influenced the mechanical stimuli experienced by cells and thereby affected bone tissue regeneration. Geometry and stiffness were controlled by changing the diameter of circular pores in the axial direction.

Despite these advancements, a comprehensive understanding of how functional grading strategies interact with scaffold material properties to influence mechanical stimulus distribution remains incomplete, particularly in clinically relevant scenarios where scaffolds are used in conjunction with rigid fixation plates. Bone scaffolds are typically fabricated from a wide range of biomaterials, spanning from soft polymers to rigid ceramics (e.g., hydroxyapatite) and metals such as titanium, each influencing biomechanical responses in distinct ways [45-48]. Consequently, it is critical to investigate how scaffold material properties affect the effectiveness of functional grading strategies.

In this study, we systematically investigate how different functional grading strategies, specifically porosity gradients, influence mechanical stimulus distribution while accounting for the effects of various biomaterials. We examine a range of materials, from low-stiffness polymers to rigid bioceramics and metals, assessing how their mechanical properties impact strain distribution. Using FEA, we evaluate the mechanical environment within large scaffolds composed of body-centred cubic (BCC) unit cells, implanted into segmental femoral defects and stabilised with a stainless-steel bone plate to replicate clinical fixation conditions. Key parameters, including the maximum-to-minimum porosity ratio, gradient resolution, scaffold material Young's modulus, and fixation plate dimensions, are systematically varied to assess their effects on osteogenic strain distribution. By integrating computational modelling with FG scaffold design, this study aims to provide quantitative insights into the interplay between scaffold architecture and material properties, paving the way for the development of next-generation scaffolds with optimised mechanical and biological performance.

## 2. Materials and Methods

### 2.1. Anatomical model preparation and defect reconstruction

The computer-aided design (CAD) model of a composite femur (Pacific Research Laboratory, Sawbones, USA) was used to develop anatomically accurate models for subsequent numerical analyses. According to the manufacturer, the femur geometry was digitised using an optical white light scanner and processed with FlexScan 3D, with surface refinements performed in Rapidform XOR3 to ensure high precision. The cortical and cancellous regions were then assembled into a final solid model in SOLIDWORKS (Dassault Systèmes, Waltham, MA, USA), preserving anatomically distinct material domains for cortical and trabecular bone.

To model scaffold implantation, a hollow cylindrical scaffold was designed with an inner diameter of 13 mm, an outer diameter of 27 mm, and a height of 30 mm, corresponding to the dimensions of the human femoral midshaft. A 30 mm segmental defect was created in the mid-diaphyseal region of the femur to accommodate the scaffold, as illustrated in Figure 1. To stabilise the defect, a 10-hole stainless steel bone plate was aligned along the femur and secured using cortical screws, replicating standard clinical fixation strategies for midshaft fractures.

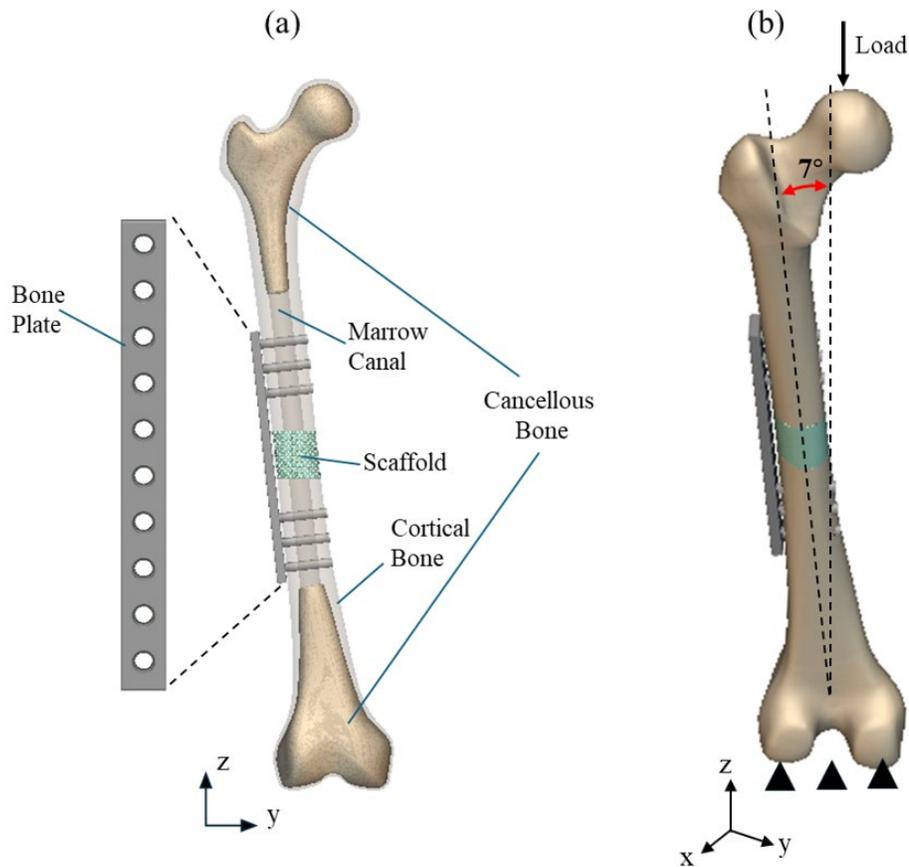

*Figure 1. CAD of a femoral segmental defect with a scaffold implant and orthopaedic fixation. The model includes cortical and cancellous bone segments, marrow canal, and a cylindrical scaffold inserted in the mid-diaphyseal region. A bone plate is attached laterally to stabilise the construct. Boundary conditions applied in the simulation include fixation of the distal femur in all directions (indicated by triangular constraints) and a physiological load applied proximally at a 7° angle to the femoral axis.*

## 2.2. Functionally graded scaffold design

The BCC lattice structure was selected as the primary architecture for the scaffold designs. To investigate the effects of porosity grading on mechanical stimulus distribution, the scaffold structures were systematically divided into multiple transverse partitions, as shown in Figure 3. These partitions were oriented perpendicular to the scaffold's central axis and maintained a fixed alignment relative to the fixation plate. The porosity progressively increased towards the plate side to reduce mechanical stimulation adjacent to the rigid fixation plate due to stress shielding [18]. This design enabled the assessment of how variations in the porosity profile influence stress shielding and mechanical stimulus distribution within the scaffold.

To parametrically control porosity in the BCC lattice unit cells, we established a quantitative relationship between strut thickness (S) and porosity (n) for unit cells with edge lengths of 1.5 mm. As shown in Figure 2, increasing the strut thickness led to a nonlinear decrease in porosity, significantly altering the internal geometry of the unit cell. A third-order polynomial was fitted to the measured data with excellent accuracy ($R^2 = 1.000$), yielding the following empirical equation.

$$n = 1.3775 \times S^3 - 2.3227 \times S^2 - 0.0359 \times S + 1.0036 \tag{1}$$

This equation was subsequently used to determine the appropriate strut thickness required to achieve target porosity levels across the scaffold, enabling systematic design of graded architectures through geometric tuning of the BCC lattice.

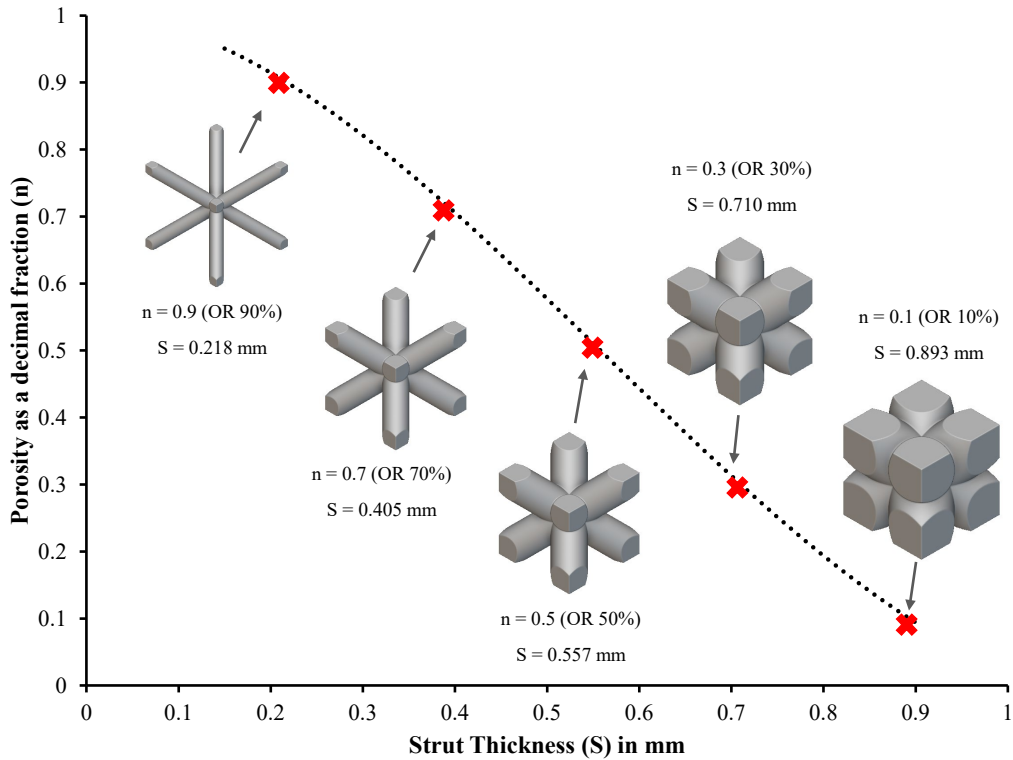

*Figure 2. Relationship between strut thickness and porosity in BCC unit cells with 1.5 mm edge length.* *A third-order polynomial fit ($R^2 = 1.000$) was used to model the nonlinear decrease in porosity (n) as strut thickness (S) increases. Representative 3D renderings of BCC unit cells are shown for selected porosity values (10%–90%), illustrating the corresponding changes in geometry.*

Scaffolds were initially developed with three-section configurations corresponding to different maximum-to-minimum porosity ratios: 50/50 (uniform), 70/50, 80/50, and 90/50. These configurations were selected to examine how the transition from low to high porosity influences both the magnitude and spatial uniformity of strain distribution within the scaffold.

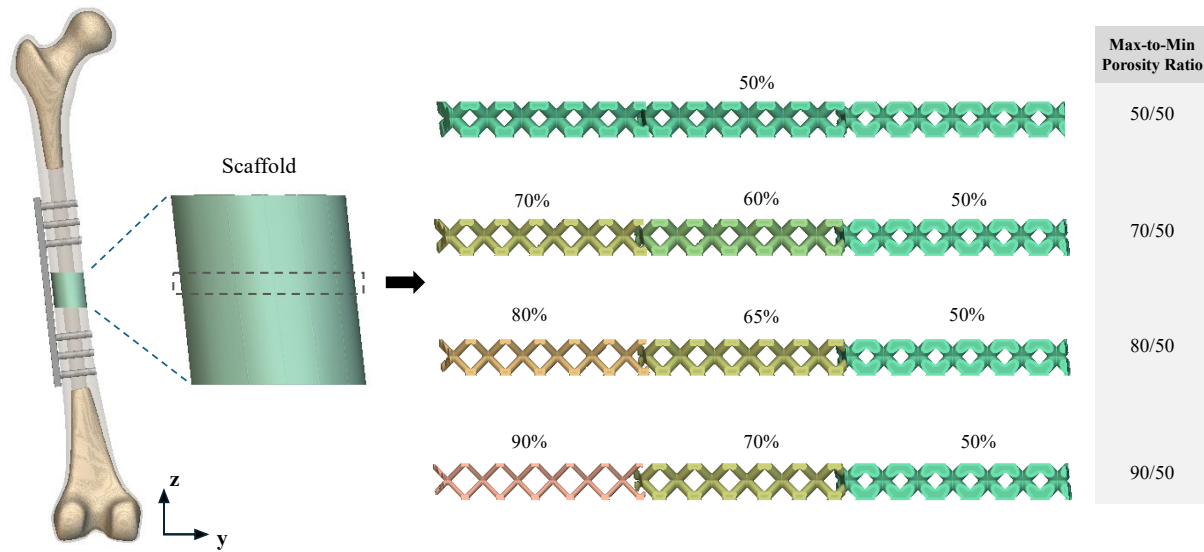

*Figure 3. Design of FG scaffolds with varying porosity profiles for load-bearing bone defect reconstruction.* A cylindrical scaffold segment is shown within the mid-diaphyseal region of a femur model. Porosity is graded along the transverse (y) direction of the scaffold, perpendicular to the bone's longitudinal axis (z-direction). Four scaffold designs are illustrated with different porosity gradients, defined by the percentage porosity at three locations across the gradient. The "Max-to-Min Porosity Ratio" represents the ratio of maximum to minimum porosity along the y-direction (e.g., 90/50 corresponds to 90% maximum and 50% minimum porosity). This metric enables direct comparison of the gradient magnitude across scaffold designs.

In addition to the three-section configurations, further scaffold models with increased gradient resolution were created, incorporating four and five section transverse partitions with evenly distributed porosity steps, as depicted in Figure 4. This approach enabled a more detailed investigation of how finer porosity transitions affect scaffold mechanics. By systematically increasing the number of partitions, the study provided a comparative evaluation of how multi-section scaffolds influence the uniformity of mechanical stimuli and the potential for more homogeneous tissue ingrowth.

All scaffold geometries were generated using nTop (nTopology Inc, New York, NY, USA), which enabled parametric control of lattice structures. The porosity gradients were created by adjusting the strut thickness within the BCC unit cells to achieve the target porosity values while maintaining the overall unit cell dimensions.

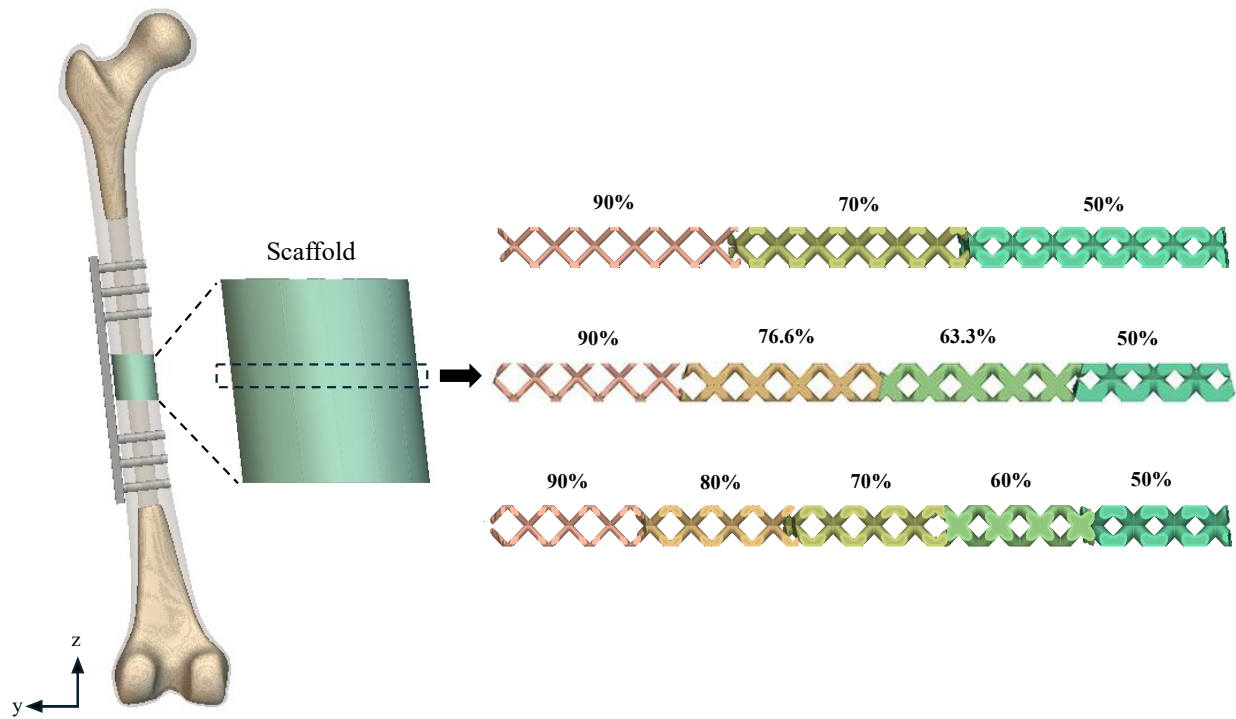

*Figure 4. FG scaffold designs with identical Max-to-Min Porosity Ratios (90/50) but different porosity gradation resolutions. The scaffolds are designed to fit within the mid-diaphyseal region of a long bone, with porosity graded along the transverse (y) direction. All three designs begin at 90% porosity and decrease to 50%, but differ in the number of intermediate porosity levels used to achieve this transition. From top to bottom, the scaffolds exhibit a low, medium, and high resolution of porosity gradation. This allows investigation of how the smoothness of the porosity transition influences mechanical performance and biological responses.*

## 2.3. Numerical simulations

FEA was conducted using ABAQUS (SIMULIA, Providence, RI, USA) to evaluate the mechanical performance of scaffolds implanted within mid-shaft segmental defects of the composite femur model, as depicted in Figure 1. A ramped compressive load with a magnitude of 1750 N (approximately 2.5 times body weight) was applied at a 7° angle relative to the femoral axis to replicate physiological loading conditions during walking [49]. The distal end of the femur was fully fixed in all degrees of freedom to simulate boundary support during weight-bearing.

The bone model, consisting of cortical and cancellous regions, was assumed to be homogeneous and isotropic. The cortical bone was assigned a Young's modulus (E) of 16.7 GPa, and the cancellous bone 155 MPa. The fixation system, including the plate and screws, was modelled as stainless steel with a Young's modulus of 180 GPa and a Poisson's ratio of 0.3. To simulate fixation, tied contact constraints were applied between the screw shafts and the pre-drilled femoral holes, as well as between the screw and the slots of the bone plate. Additionally, a tied contact was used between the scaffold surface and the surrounding bone to ensure stable scaffold integration.

The material forming the scaffold was assumed to be isotropic, with Young's modulus values ranging from 1 GPa to 200 GPa to simulate different biomaterial conditions. To reduce computational cost, the detailed lattice architectures were replaced with homogenised solid models with equivalent effective material properties, as shown in Figure 5. All scaffold models were analysed using identical simulation setups to enable direct comparisons of mechanical performance across different porosity gradients and material properties. To ensure modelling accuracy while maintaining computational efficiency, a mesh convergence analysis was conducted using 4-node linear tetrahedral elements (C3D4).

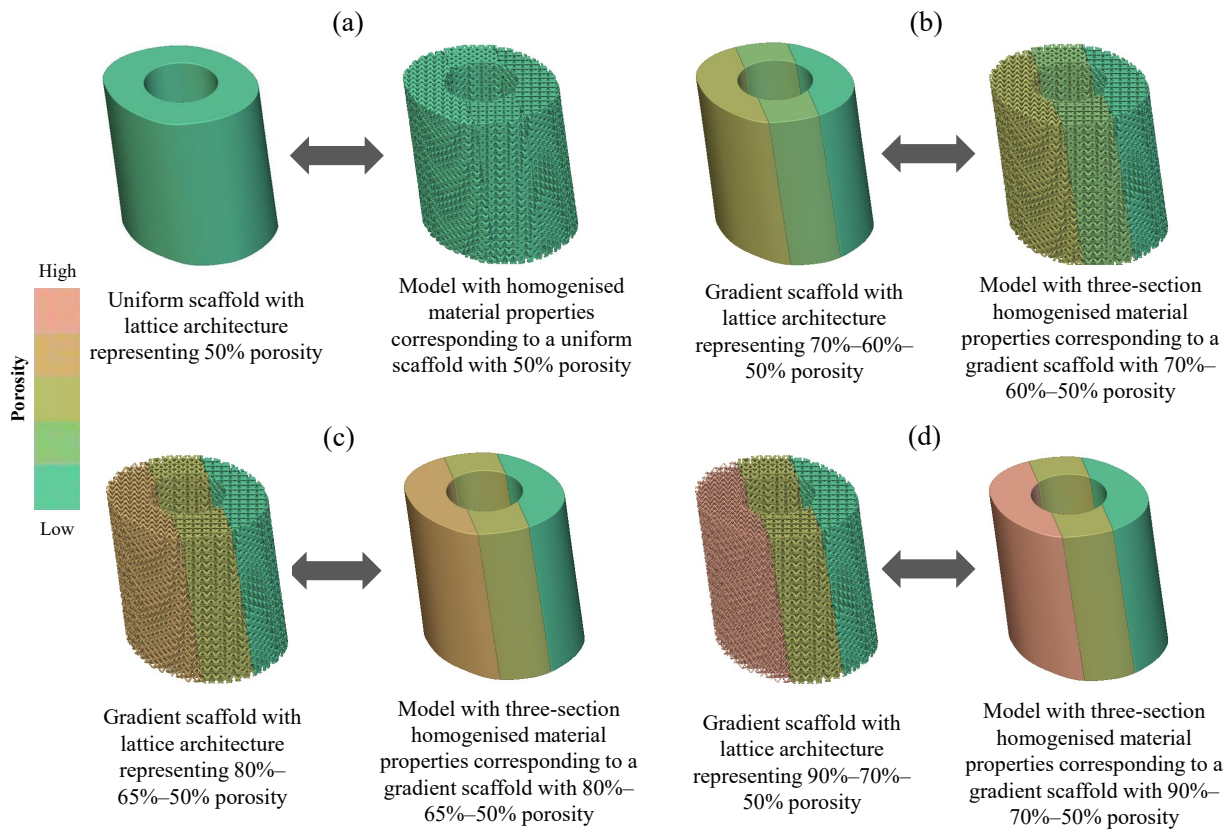

*Figure 5. Comparison between lattice-based and homogenised scaffold models with different porosity gradient profiles. Each pair of images shows a scaffold with explicit lattice architecture (left) and its corresponding homogenised solid model (right), divided into three regions with equivalent effective material properties. Four porosity profiles are illustrated: (a) uniform 50%, (b) graded 50%–60%–70%, (c) graded 50%–65%–80%, and (d) graded 50%–70%–90%. The homogenised models simplify the geometry while preserving the spatial variation in porosity-dependent stiffness, enabling more computationally efficient simulations.*

The efficiency and accuracy of this homogenisation approach were validated against experimental digital image correlation (DIC) measurements in our previous study [23]. As illustrated in Figure 6, composite femur models implanted with porous scaffolds were subjected to uniaxial compressive loading in a mechanical testing setup, and surface deformations were captured using a stereo DIC system. The femur was vertically loaded through the femoral head, simulating physiological compression while the scaffold and lateral fixation plate stabilised the segmental defect. The DIC system enabled full-field 3D deformation measurements on the scaffold surface, which were directly compared with deformation results predicted by finite element (FE) simulations using the homogenised scaffold models. The close agreement between experimental and numerical deformation fields validated the homogenisation strategy, demonstrating its capability to reliably capture scaffold mechanical behaviour while significantly reducing computational cost.

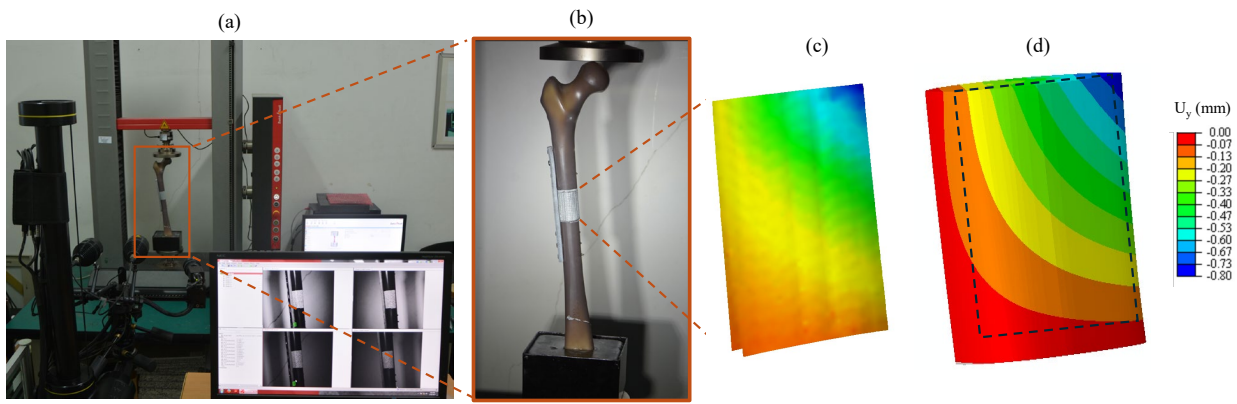

*Figure 6. Validation of the homogenised scaffold modelling approach using experimental digital image correlation (DIC) from our previous study [23]. (a) DIC-enabled mechanical testing setup used to capture femur surface deformation under uniaxial compression. (b) Close-up of the femur with implanted scaffold and applied speckle pattern. (c) Experimental vertical displacement (Uy) map obtained via DIC. (d) Corresponding Uy displacement field predicted by finite element simulation of the homogenised model. The strong agreement confirms the accuracy of the homogenisation approach.*

## 2.4. Homogenisation method

To reduce the computational cost of the FEA simulations, the porous structures were substituted with homogenised solid models of comparable size, as shown in Figure 5. Each homogenised model was assigned effective moduli matching those of its corresponding porous scaffold to ensure mechanical equivalence [23]. To capture directional variations in stiffness, the homogenised blocks were modelled with orthotropic material properties, where their linear elastic behaviour was defined using nine independent elastic stiffness constants (in elastic tensor D) as follows:

$$\begin{Bmatrix} \sigma_{11} \\ \sigma_{22} \\ \sigma_{33} \\ \sigma_{12} \\ \sigma_{13} \\ \sigma_{23} \end{Bmatrix} = \begin{bmatrix} D_{1111} & D_{1122} & D_{1133} & 0 & 0 & 0 \\ & D_{2222} & D_{2233} & 0 & 0 & 0 \\ & & D_{3333} & 0 & 0 & 0 \\ & sym & & D_{1212} & 0 & 0 \\ & & & & D_{1313} & 0 \\ & & & & & D_{2323} \end{bmatrix} \begin{Bmatrix} \varepsilon_{11} \\ \varepsilon_{22} \\ \varepsilon_{33} \\ \varepsilon_{12} \\ \varepsilon_{13} \\ \varepsilon_{23} \end{Bmatrix} \quad (2)$$

These 9 elasticity constants that define the D tensor were comprised of the 3 directional Young's moduli $E_{11}, E_{22}, E_{33}$ and 3 shear moduli $G_{12}, G_{13}, G_{23}$ as below:

$$D_{1111} = E_{11}(1 - \nu_{23}\nu_{32})\Upsilon \quad (3)$$

$$D_{2222} = E_{22}(1 - \nu_{13}\nu_{31})\Upsilon \quad (4)$$

$$D_{3333} = E_{33}(1 - \nu_{12}\nu_{21})\Upsilon \quad (5)$$

$$D_{1122} = E_{11}(\nu_{21} - \nu_{31}\nu_{23})\Upsilon = E_{22}(\nu_{12} - \nu_{32}\nu_{13})\Upsilon \quad (6)$$

$$D_{1133} = E_{11}(\nu_{31} - \nu_{21}\nu_{32})\Upsilon = E_{33}(\nu_{13} - \nu_{12}\nu_{23})\Upsilon \quad (7)$$

$$D_{2233} = E_{22}(\nu_{32} - \nu_{12}\nu_{31})\Upsilon = E_{33}(\nu_{23} - \nu_{21}\nu_{13})\Upsilon \quad (8)$$

$$D_{1212} = G_{12} \quad (9)$$

$$D_{1313} = G_{13} \quad (10)$$

$$D_{2323} = G_{23} \quad (11)$$

where;

$$\Upsilon = \frac{1}{1 - \nu_{12}\nu_{21} - \nu_{23}\nu_{32} - \nu_{31}\nu_{13} - 2\nu_{21}\nu_{32}\nu_{13}} \quad (12)$$

The directional Young's moduli ($E_{11}, E_{22}, E_{33}$) and shear moduli ($G_{12}, G_{13}, G_{23}$) were calculated using an asymptotic homogenisation approach with periodic boundary conditions, as detailed in [23, 50].

## 2.5. Problem Definition

The objective of this study was to optimise the porosity distribution within an FG scaffold to enhance mechanical stimuli conducive to bone regeneration, while minimising variation in the mechanical environment. Specifically, we aimed to maximise the mean octahedral shear strain within the scaffold to promote osteogenic stimulation, while simultaneously minimising the coefficient of variation (CV), defined as the ratio of the standard deviation to the mean, to achieve a more uniform strain distribution and, consequently, more homogeneous bone formation.

The mathematical formulation of the objective can be expressed as:

$$\max(\bar{\varepsilon}_{oct}), \quad \min(\sigma_{\varepsilon_{oct}} / \bar{\varepsilon}_{oct}) \tag{13}$$

where $\bar{\varepsilon}_{oct}$ and $\sigma_{\varepsilon_{oct}}$ denote the mean and standard deviation of octahedral shear strain, respectively.

Mathematically, the mean and standard deviation of element octahedral shear strain are given by:

$$\bar{\varepsilon}_{oct} = \frac{\sum_{i=1}^{N} \varepsilon_{oct,i} V_i}{\sum_{i=1}^{N} V_i}, \quad \sigma_{\varepsilon_{oct}} = \sqrt{\frac{\sum_{i=1}^{N} V_i (\varepsilon_{oct,i} - \bar{\varepsilon}_{oct})^2}{\sum_{i=1}^{N} V_i}} \tag{14}$$

where $i$ is the element index, $\varepsilon_{oct,i}$ is the octahedral shear strain in element $i$, $V_i$ is the element volume and N is the total number of elements in the FG scaffolds.

Octahedral shear strain was selected as the mechanical stimulus parameter because it captures the combined effects of multiaxial deformation modes (tension, compression, and shear) and has been widely correlated with mechano-regulation and tissue differentiation in bone regeneration studies [51].

The octahedral shear strain was calculated based on the principal strains as:

$$\varepsilon_{oct} = \frac{2}{3} \sqrt{(\epsilon_1 - \epsilon_2)^2 + (\epsilon_2 - \epsilon_3)^2 + (\epsilon_3 - \epsilon_1)^2} \tag{15}$$

where $\epsilon_1$, $\epsilon_2$, and $\epsilon_3$ are the principal normal strains.

By reducing the CV in octahedral shear strain, the strain distribution within the scaffold becomes more uniform, minimising regions with low mechanical stimulation that could otherwise impair bone ingrowth. Simultaneously, maintaining a high mean strain level ensures that the scaffold experiences sufficient mechanical cues to promote an osteoconductive environment conducive to tissue regeneration. These design principles align with established mechanobiological models of bone remodelling, where strain-mediated stimulation plays a critical role in guiding cellular responses [52].

It should be noted that the approach adopted in this study did not involve an automated optimisation algorithm. Instead, the porosity gradient profiles were manually adjusted for different material types, and their effects were evaluated based on strain distribution metrics. The defined objective function served as a guiding principle rather than being computationally solved through a formal optimisation procedure.

## 3. Results and Discussion

The porosity grading strategy adopted in this study enabled scaffold porosity to progressively increase toward the side adjacent to the bone plate, with the aim of enhancing load transfer and mechanical stimulation in regions typically affected by stress shielding. The rationale for increasing porosity towards the plate side was based on previous studies reporting reduced mechanical stimulation adjacent to fixation plates due to stress shielding [18]. The BCC lattice structure was selected as the scaffold architecture due to its widely recognised mechanical efficiency and demonstrated suitability for biomedical applications [53]. From a biological perspective, the open-cell nature of BCC lattices enhances permeability for fluid transport, nutrient diffusion, and cell migration. The interconnected porous network further provides an ideal environment for cell adhesion and proliferation, supporting bone tissue regeneration [54, 55].

To evaluate the effectiveness of this approach, FEA was conducted to compare three different porosity gradient configurations against a corresponding uniform scaffold, as illustrated in Figure 3. All scaffold models were assigned homogenised equivalent material properties to enable accurate mechanical evaluation while maintaining computational efficiency. The accuracy of this homogenisation technique was previously validated in our experimental study using DIC to measure deformation and strain distributions in scaffolds implanted within segmental defects in Sawbone femur composite bones [23].

Octahedral shear strain was used for evaluating scaffold performance, as it is widely recognised for its ability to represent local distortional deformation relevant to osteogenic activity. Strain was measured at the scaffold surface, reflecting the early mechanical environment experienced by cells immediately after implantation, when cell attachment is predominantly limited to the pore surfaces. This assumption is supported by recent findings suggesting that surface strains, rather than volumetric strains, are more appropriate indicators of the initial mechanical cues governing cellular behaviour during early-stage bone regeneration [51, 56].

Figure 7a illustrates the percentage change in mean octahedral shear strain within the scaffold for each gradient configuration, relative to the uniform scaffold. A positive percentage change indicates an enhancement of mechanical stimuli compared to the uniform scaffold. Figure 7b presents the corresponding percentage change in the CV of octahedral shear strain, with a more negative value representing improved strain homogeneity. Each group of bars represents a different porosity gradient pattern (50%–60%–70%, 50%–65%–80%, and 50%–70%–90%), while the bar colour intensity corresponds to the scaffold material's Young's modulus, ranging from 1 GPa to 200 GPa, as indicated in the legend.

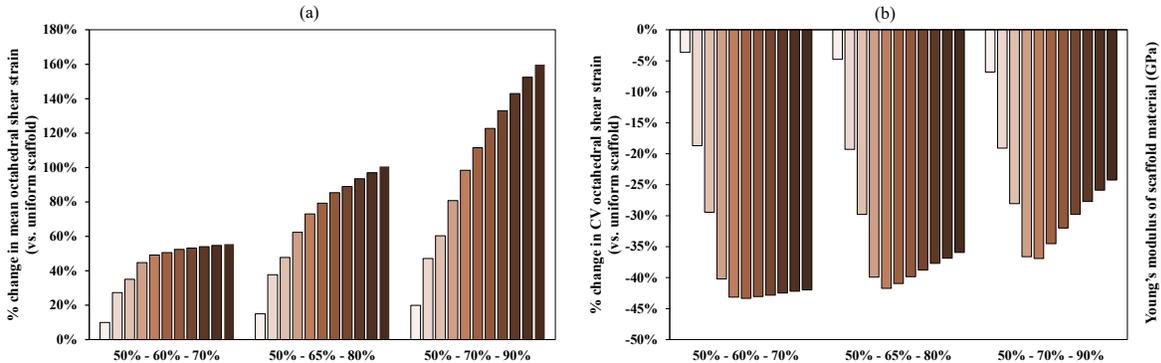

*Figure 7. Effect of porosity gradient magnitude and scaffold material stiffness on the octahedral shear strain distribution within the scaffold region. (a) Percentage change in mean octahedral shear strain within the scaffold region, relative to the corresponding uniform scaffold. (b) Percentage change in the CV of octahedral shear strain within the scaffold region, also relative to the uniform scaffold. CV is calculated as the ratio of the standard deviation to the mean strain, providing a measure of strain heterogeneity. Each group of bars represents a different porosity distribution pattern (e.g., 50%–60%–70%), and bar colour intensity corresponds to scaffold material Young's modulus (1–200 GPa), as shown in the grayscale legend.*

Across all scaffold materials, introducing a porosity gradient markedly enhanced the mechanical stimulus compared to the uniform scaffold, as indicated by the positive changes in mean octahedral shear strain (Figure 7a). However, the magnitude of this improvement varied depending on Young's modulus of the scaffold material. Scaffolds with intermediate-to-high stiffness values (25–200 GPa), representative of ceramic and metallic biomaterials, exhibited the greatest enhancement in mean strain with increasing gradient magnitude. In contrast, scaffolds with low stiffness values (1–10 GPa), typical of polymeric materials, showed comparatively smaller improvements, suggesting that gradient strategies are more effective in scaffolds capable of sustaining and redistributing higher mechanical loads.

The reduction in strain heterogeneity, reflected by the decrease in the CV (Figure 7b), was more pronounced in scaffolds with moderate-to-high Young's moduli. This indicates that functional grading not only amplifies mechanical stimulation but also promotes more uniform strain distribution when the scaffold possesses sufficient intrinsic stiffness. However, increasing the max-to-min porosity ratio does not consistently enhance strain homogeneity, and in some cases, particularly for stiffer materials, the CV slightly worsens. Nevertheless, when both mean strain and CV are considered together, the overall mechanical stimulation outcome still improves with increasing gradient magnitude. For instance, at a Young's modulus of 100 GPa, the mean strain improvement increases substantially from 52% to 123%, while the reduction in CV diminishes only slightly from 43% to 32%. These findings demonstrate that increasing the porosity gradient magnitude remains effective, even when minor trade-offs in strain homogeneity occur.

To further illustrate the effect of porosity gradient magnitude on mechanical stimulus distribution, a representative scaffold composed of titanium alloy (Ti-6Al-4V) was selected, given its existing use in clinical applications for load-bearing orthopaedic implants. Figure 8 shows the distribution of octahedral shear strain within the scaffold for different porosity gradient magnitudes. The scaffold models are longitudinally sectioned to reveal the internal strain patterns along the transverse (y) direction, where the porosity gradient was applied.

The uniform scaffold with 50% porosity (Figure 8a) exhibited relatively low octahedral shear strain values, particularly adjacent to the fixation plate, indicating regions of stress shielding. In contrast, the FG scaffolds (Figures 8b–d) demonstrated progressively higher strain levels and more extensive strain distribution within the scaffold as the maximum-to-minimum porosity ratio increased. Specifically, the scaffold with the 90%–70%–50% gradient (Figure 8d) exhibited the highest strain concentrations in regions near the plate, suggesting that greater porosity variation enhances load transfer into the scaffold and mitigates stress shielding effects. These visual trends are consistent with the quantitative findings presented in Figure 7, highlighting the potential of porosity grading to improve mechanical stimulation within scaffolds implanted in segmental bone defects.

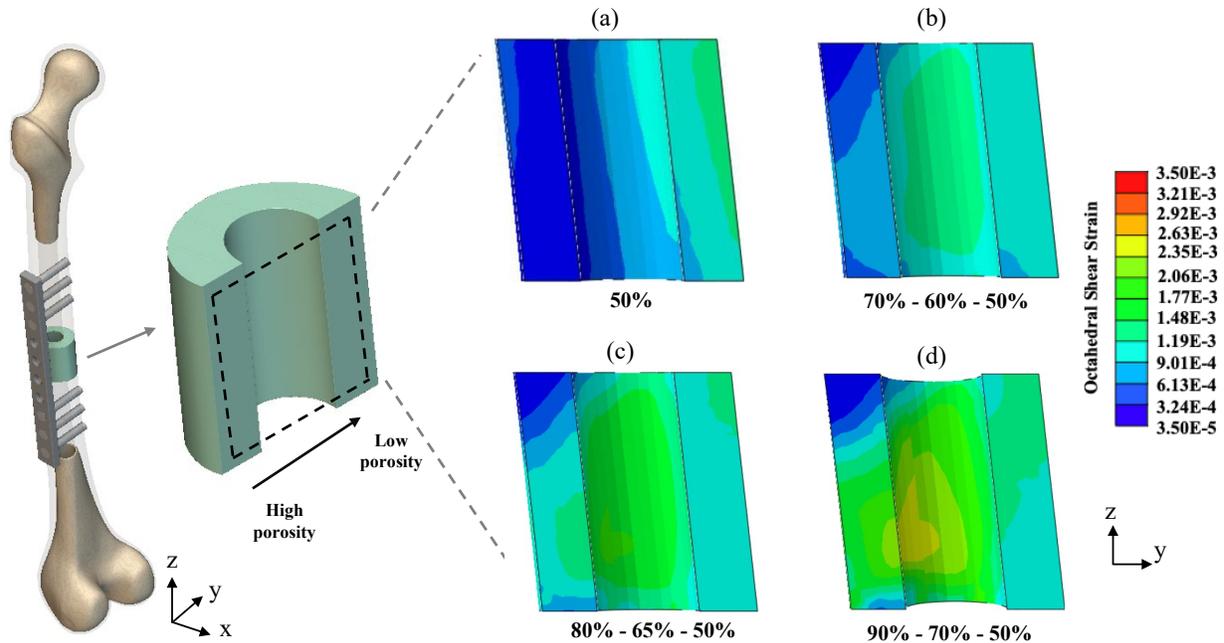

***Figure 8. Distribution of octahedral shear strain within a titanium scaffold for various porosity gradient magnitudes.*** *The cylindrical scaffold is assumed to be composed of Ti-6Al-4V and is longitudinally sectioned to reveal internal strain distribution along the transverse (y) direction. (a) Uniform scaffold with 50% porosity. (b–d) FG scaffolds with increasing porosity gradient magnitude: (b) 50%–60%–70%, (c) 50%–65%–80%, and (d) 50%–70%–90%. Contour plots show octahedral shear strain distributions (unitless) under physiological compressive loading.*

Following the evaluation of porosity gradient magnitude, we further examined how the smoothness of the porosity transition, referred to as the gradient resolution, influences the mechanical environment within the scaffold. Figure 9 presents the results for scaffolds with identical maximum-to-minimum porosity ratios (90/50) but differing numbers of intermediate porosity steps. The three gradient resolutions evaluated were: (i) a three-section gradient (50%–70%–90%), (ii) a four-section gradient (50%–63.3%–76.6%–90%), and (iii) a five-section gradient (50%–60%–70%–80%–90%).

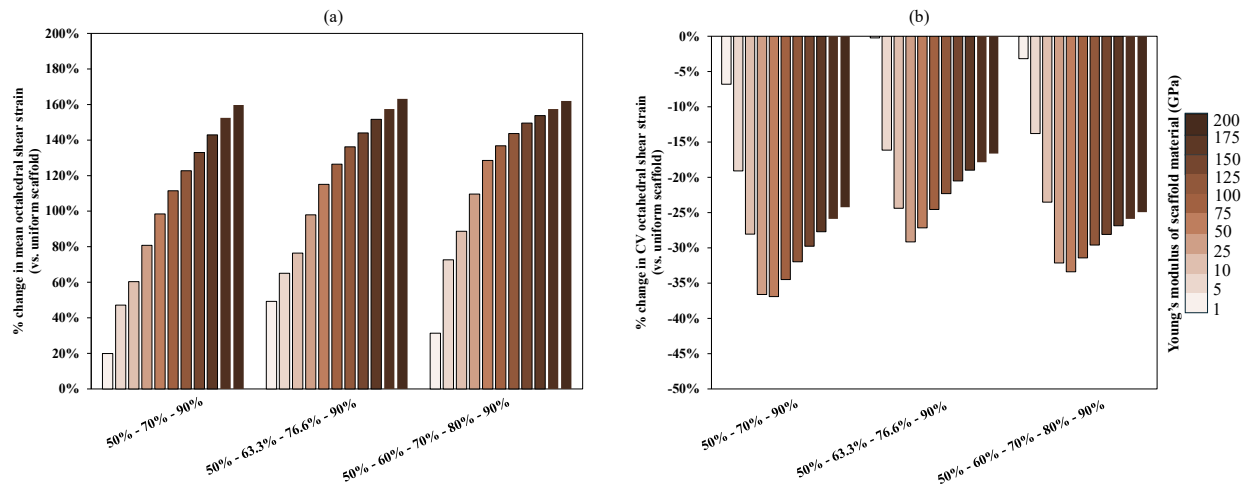

*Figure 9. Effect of porosity gradient resolution and scaffold material stiffness on the mechanical strain environment. (a) Percentage change in mean octahedral shear strain within the scaffold region, relative to the corresponding uniform scaffold. (b) Percentage change in the CV of octahedral shear strain within the scaffold region, also relative to the uniform scaffold. CV is calculated as the ratio of the standard deviation to the mean strain, providing a measure of strain heterogeneity. Each group of bars represents a different porosity gradient resolution—50%–70%–70%, 50%–63.3%–76.6%–90%, and 50%–60%–70%–80%–90%—with bar shading indicating scaffold material Young's modulus (1–200 GPa), as shown in the grayscale legend.*

As shown in Figure 9a, increasing the gradient resolution consistently enhanced the mean octahedral shear strain within the scaffolds. These improvements were observed across the entire range of scaffold Young's moduli (1–200 GPa), although the relative increases in mean strain with higher gradient resolution were more pronounced for scaffolds with lower and intermediate stiffness values (1–100 GPa). This may suggest that substantial relative improvements were already achieved in stiffer scaffold materials through the application of porosity grading (as illustrated in Figure 7), whereas the increase in gradient resolution could be particularly helpful for further enhancing mechanical stimuli in scaffolds composed of lower stiffness materials. The observed improvements can be attributed to the fact that gradual transitions help maintain consistent load transfer and mechanical stimulation across the scaffold, thereby reducing regions of mechanical under-stimulation.

While increasing the gradient resolution enhanced the mean octahedral shear strain within the scaffold, its effect on strain homogeneity was less favourable (Figure 9b). In fact, the CV slightly worsened across most material stiffness ranges as the gradient resolution increased. Nevertheless, the overall outcome can still be considered positive, as the improvement in mean strain was substantially greater than the minor compromise in CV. For instance, in a material with a Young's modulus of 100 GPa, increasing the gradient resolution from three to five sections improved the percentage change in mean strain (relative to the uniform scaffold) from 123% to 144%, while the percentage change in coefficient of variation (CV) was diminished by only 2%.

To visually demonstrate the influence of gradient resolution on mechanical stimuli distribution, representative titanium scaffold models corresponding to the different gradient strategies analysed in Figure 9 were longitudinally sectioned and evaluated, as shown in Figure 10. Comparing the FG designs, the three-section scaffold (90%–70%–50%, Figure 10b) demonstrated a notable increase in strain magnitude and spatial extent compared to the uniform scaffold. Further increasing the gradient resolution to four sections (90%–75%–60%–50%, Figure 10c) and five sections (90%–80%–70%–60%–50%, Figure 10d) enhanced the coverage of high strain regions within the scaffold. These visual findings agree with the quantitative improvements observed with higher gradient resolution and underscore the importance of smooth porosity transitions for optimising scaffold mechanical performance.

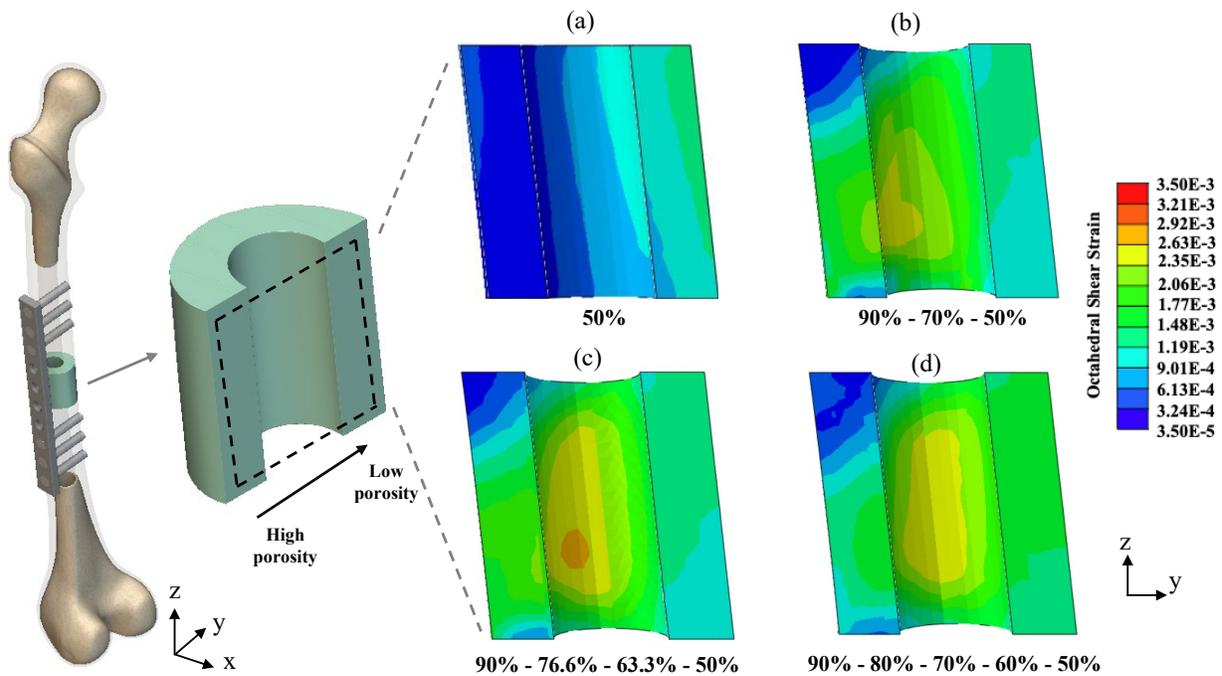

*Figure 10. Octahedral shear strain distribution within a titanium scaffold for uniform and FG scaffold designs with increasing gradient resolutions.* The cylindrical scaffold is longitudinally sectioned to reveal internal strain distribution. (a) Uniform scaffold with 50% porosity. (b–d) FG scaffolds with the same max-to-min porosity ratio (90% to 50%) but increasing gradient resolution: (b) 90%–70%–50%, (c) 90%–75%–60%–50%, and (d) 90%–80%–70%–60%–50%. Contour plots display octahedral shear strain (unitless) under physiological loading.

To further evaluate factors influencing the mechanical environment within the scaffold, we investigated the effect of fixation rigidity by varying plate dimensions, specifically plate width and thickness, on strain distribution. The plate length and the number and positions of screws were kept constant, while three configurations with distinct cross-sectional geometries were analysed to capture differences in rigidity. As shown in Figure 11, the three plates differ only in their width and thickness. Plate No. 2, which was used in all previous simulations and results, served as the reference design. Plates No. 1 and No. 3 were used to examine whether variations in plate rigidity, due to changes in cross-sectional geometry, would significantly affect the mean and coefficient of variation (CV) of octahedral shear strain within the scaffold.

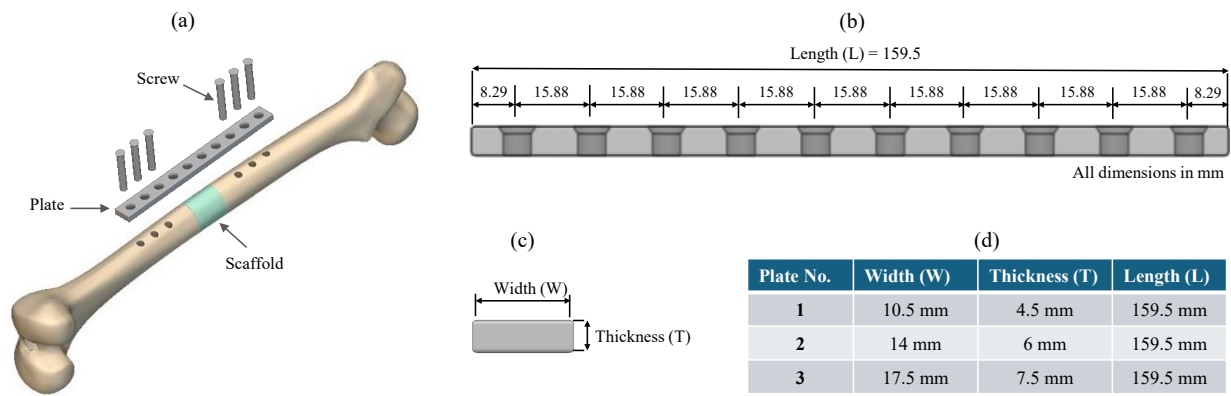

*Figure 11. Fixation plate configurations used to assess the effect of plate stiffness on scaffold strain distribution.* *(a) Schematic of the femur model with scaffold and plate fixed using cortical screws. (b) Top view showing the screw positions and constant plate length (L = 159.5 mm) across all configurations. (c) Cross-sectional schematic defining plate width (W) and thickness (T), the two parameters varied in this analysis. (d) Table summarising the dimensions of the three plate designs.*

To assess how the fixation plate rigidity influences the effectiveness of functional grading, we analysed how plate stiffness affects the relative improvement in mean strain and the change in strain heterogeneity (CV) when switching from a uniform scaffold to a functionally graded scaffold. As shown in Figure 12, the increase in mean octahedral shear strain (left) resulting from porosity grading was more pronounced in configurations with lower plate stiffness (i.e., Plate 1), whereas stiffer plates (i.e., Plate 3) showed reduced improvement. While the mean strain showed a consistent reduction in improvement with increasing plate stiffness, the trend in strain heterogeneity was not strictly linear. Specifically, the change in CV was more substantial when transitioning from Plate 1 to Plate 2 but showed minimal further reduction and even slight reversal for Plate 3. This nonlinearity may be attributed to complex interactions between plate stiffness and local scaffold deformation. As the plate becomes significantly stiffer, it may dominate the load transfer path and suppress strain magnitudes uniformly, but without further improving strain uniformity. In such cases, the scaffold region near the plate may remain under-stimulated, leading to persistent heterogeneity in strain distribution. These findings highlight that beyond a certain threshold, increasing fixation stiffness may no longer yield proportional gains in mechanical uniformity, and may even plateau or reverse its effects on scaffold mechanobiology.

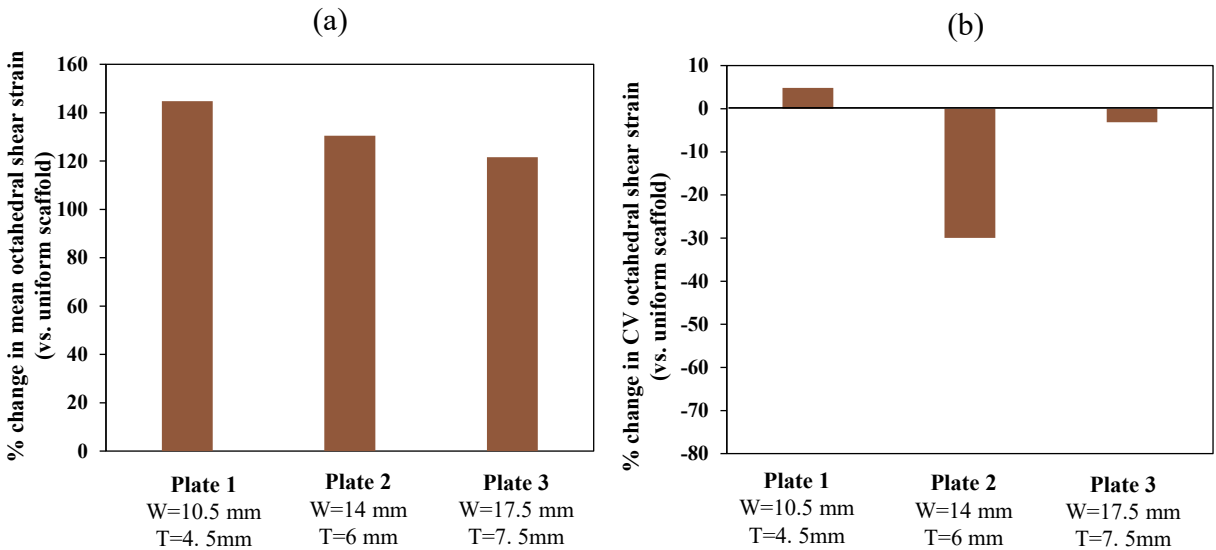

*Figure 12. Influence of plate geometry on the effectiveness of functional grading. (a) Percentage increase in mean octahedral shear strain within the scaffold when switching from a uniform to a functionally graded (FG) scaffold. (b) Change in strain heterogeneity (CV) between the uniform and FG scaffold. Results are shown for all three plate types. Softer plates (e.g., Plate 1) showed greater improvements from functional grading, suggesting plate stiffness modulates the mechanical benefits of scaffold design strategies.*

To consolidate findings for the mechanical benefits of functional grading, a direct quantitative comparison was conducted between a uniform scaffold and a functionally graded scaffold incorporating five transverse porosity regions (50%–60%–70%–80%–90%). This comparison focused on the absolute magnitudes of mean octahedral shear strain and strain heterogeneity (CV), as shown in Figure 13. In addition to visualising the effect of porosity grading, this comparison aimed to assess the extent of mechanical improvements achieved across scaffolds fabricated from different common biomaterials. Scaffolds composed of titanium alloy (Ti-6Al-4V) (E = 113.8 GPa), bioactive glass (45S5 Bio-glass) (E = 35 GPa), and polylactic acid (PLA) (E = 2.6 GPa) were evaluated. The top row of Figure 13a presents the uniform scaffold, while the bottom row illustrates the graded scaffold, visualised through cross-sectional, side, and isometric views.

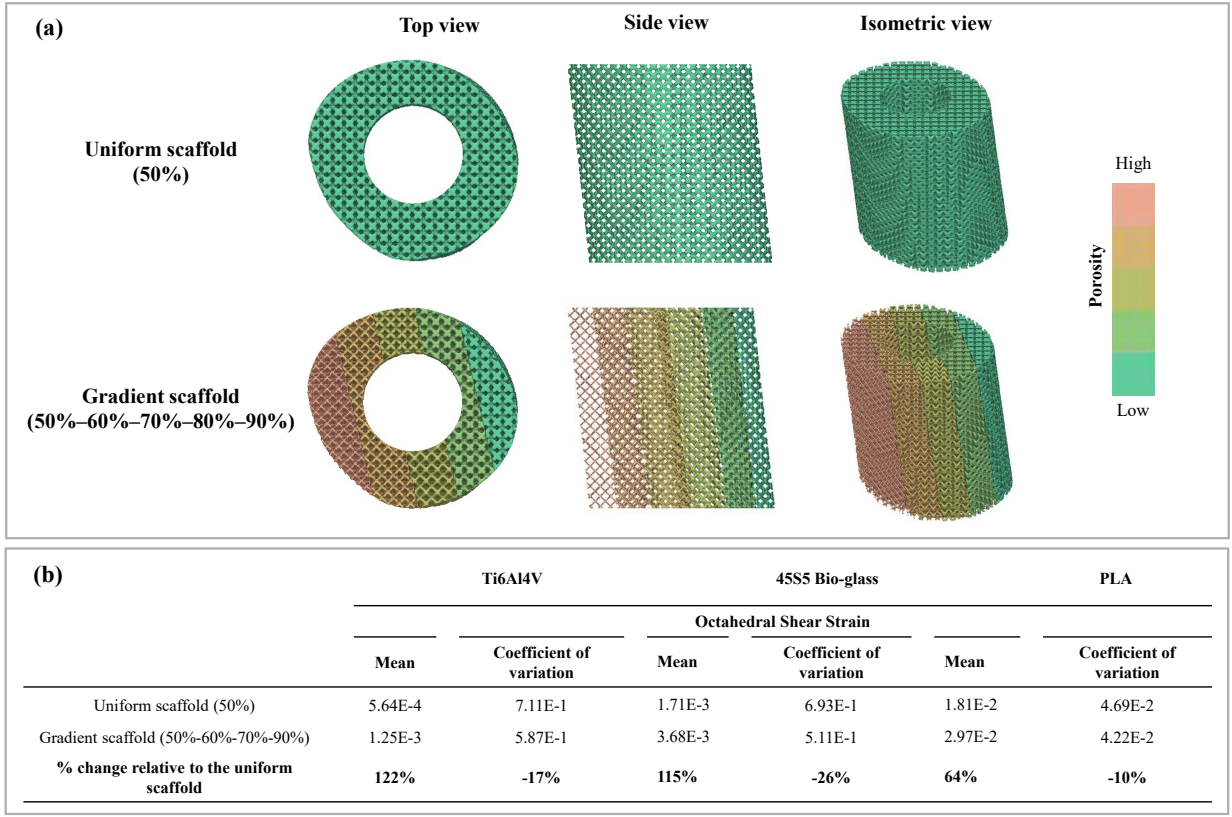

*Figure 13.* (a) Comparison between a uniform and a FG lattice scaffold. The top row shows a scaffold with uniform 50% porosity, visualised in cross-sectional, side, and full 3D views. The bottom row presents a FG scaffold with five transverse porosity regions: 50%, 60%, 70%, 80%, and 90%. Cross-sectional, side, and 3D views illustrate the spatial variation in lattice density. (b) Table summarising the absolute mean and coefficient of variation (CV) of octahedral shear strain for the uniform and FG scaffolds composed of three different materials: Ti-6Al-4V, 45S5 Bio-glass, and PLA.

The corresponding mechanical performance, summarised in Figure 13b, reinforces the substantial benefits of functional grading across different scaffold materials. For the Ti-6Al-4V scaffold, the graded design resulted in a 122% increase in mean strain and a 17% reduction in strain heterogeneity (measured by the coefficient of variation, CV) compared to the uniform scaffold. Similar trends were observed for functionally graded scaffolds composed of 45S5 Bio-glass and PLA, although with lower magnitudes of improvement relative to the uniform scaffold, reflecting the interplay between material stiffness and the effectiveness of porosity grading. Notably, in softer materials such as PLA, the absolute mean strain was already significantly higher than in stiffer materials like Ti-6Al-4V ($4.69 \times 10^{-2}$ vs. $5.64 \times 10^{-4}$). This difference may help explain why functionally graded scaffolds composed of stiffer materials exhibit a much higher percentage increase in mean strain (relative to the uniform scaffold), since they start from a lower baseline and have greater potential for mechanical enhancement.

These results validate the earlier computational findings (Figures 5–8) by demonstrating that, regardless of material selection, introducing gradual porosity transitions can significantly improve the mechanical environment within scaffolds. They further highlight that the degree of improvement is influenced by the scaffold material's intrinsic stiffness, with stiffer materials benefiting more markedly from gradient strategies.

While this study provides valuable insights into the mechanical optimisation of FG scaffolds, several limitations should be acknowledged. First, the mechanical environment was evaluated under simplified loading conditions representing static walking loads; however, bone healing *in vivo* is influenced by a broader range of dynamic and physiological load variations. Furthermore, this study focused on mechanical stimuli during the early stage of bone healing, therefore, biological factors such as cell migration, tissue ingrowth, and scaffold degradation were not incorporated into the current modelling framework. Future work should focus on developing multiscale modelling approaches that couple mechanical and biological phenomena, exploring dynamic loading scenarios, and investigating scaffold degradation [57, 58]. Additionally, experimental validation using *in vitro* or *in vivo* models will also be critical to further substantiate the computational findings and facilitate the clinical translation of the optimised scaffold designs.

## 4. Conclusion

This study systematically investigated the influence of porosity gradient magnitude, gradient resolution, and scaffold material properties on the distribution of mechanical stimuli within functionally graded (FG) bone scaffolds. By progressively increasing porosity toward the fixation plate, functional grading strategies were shown to enhance both the magnitude and uniformity of mechanical stimulation, particularly in regions prone to stress shielding. These improvements were more pronounced in scaffolds composed of materials with higher Young's modulus (>25 GPa), such as Ti-6Al-4V, highlighting the critical interplay between material selection and architectural design. Increasing the gradient resolution by introducing additional intermediate porosity steps further amplified the strain levels within the scaffold. The introduction of a five-step porosity gradient (50%–60%–70%–80%–90%) in Ti-6Al-4V scaffolds resulted in a 122% increase in mean octahedral shear strain and a 17% reduction in strain heterogeneity compared to the uniform design. Additional analysis showed that the fixation plate rigidity also modulates the mechanical benefits of functional grading, with non-linear effects observed in strain heterogeneity. These findings underscore the importance of simultaneously optimising porosity profiles and material properties to design scaffolds that more effectively promote bone regeneration. Future work incorporating dynamic loading conditions, biological responses, and experimental validation will be essential to further refine these strategies and facilitate their clinical translation.

**Ethics statement**

The authors declare that an ethics statement was not required for this study, as there was no involvement of human participants or animals.

**Declaration of generative AI and AI-assisted technologies in the writing process**

During the preparation of this work the authors used ChatGPT (OpenAI) to assist with improving the clarity and language of the text. After using this tool/service, the authors reviewed and edited the content as needed and take full responsibility for the content of the publication.

**Data availability**

Data will be made available upon request.

**Conflicts of interest**

There are no conflicts to declare.